\documentclass[aps,reprint,groupedaddress,longbibliography,nofootinbib]{revtex4-1}

\usepackage{graphicx}
\usepackage{hyperref}
\usepackage{amsmath}
\usepackage{amssymb}
\usepackage[squaren, Gray, cdot]{SIunits}
\usepackage{bm}
\usepackage[english]{babel}
\usepackage{float}
\addto\captionsenglish{}

\begin{document}


\title{Geometry dependence of surface lattice resonances in plasmonic nanoparticle arrays}


\author{R. Guo}
\author{T.K. Hakala}
\author{P. T{\"o}rm{\"a}}
\email[]{paivi.torma@aalto.fi}
\affiliation{COMP Centre of Excellence, Department of Applied Physics, Aalto University, P.O.Box 15100, FI-00076 Aalto, Finland}


\date{\today}

\begin{abstract}

Plasmonic nanoarrays which support collective surface lattice resonances (SLRs) have become an exciting frontier in plasmonics. Compared with the localized surface plasmon resonance (LSPR) in individual particles, these collective modes have appealing advantages such as angle-dependent dispersions and much narrower linewidths. Here, we investigate systematically how the geometry of the lattice affects the SLRs supported by metallic nanoparticles. We present a general theoretical framework from which the various SLR modes of a given geometry can be straightforwardly obtained by a simple comparison of the diffractive order (DO) vectors and orientation of the nanoparticle dipole given by the polarization of the incident field. Our experimental measurements show that while square, hexagonal, rectangular, honeycomb and Lieb lattice arrays have similar spectra near the $\Gamma$-point ($k=0$), they have remarkably different SLR dispersions. Furthermore, their dispersions are highly dependent on the polarization. Numerical simulations are performed to elucidate the field profiles of the different modes. Our findings extend the diversity of SLRs in plasmonic nanoparticle arrays, and the theoretical framework provides a simple model for interpreting the SLRs features, and vice versa, for designing the geometrical patterns.
\end{abstract}


\maketitle

\section{Introduction}

The conduction electron oscillations within a metallic nanoparticle driven by an external electromagnetic field gives rise to a localized surface plasmon resonance (LSPR). At the resonance, a metallic particle will confine light at the nanoscale, with the electric field being enhanced in the near-field region at the surface of the particle. The field enhancement and sub-wavelength character of LSPRs can be applied to modify the spontaneous emission decay rate of nanoemitters \cite{Anger2006} and to control various nonlinear effects, such as second harmonic generation and Raman scattering \cite{Willets2007,Hubert2007}. However, due to the strong radiative damping, LSPRs usually exhibit broad spectral linewidths and low quality factors \cite{Wang2006} which hinder potential applications. If the nanoparticles are placed in an array, the dipolar interactions between the particles may induce extra resonances. In particular, when the array periodicity is on the order of particle resonance wavelength, the coupling between the diffractive orders (DOs) of the array and the LSPRs on each individual particle will result in a collective resonance called surface lattice resonance (SLR) \cite{Zou2005,Kravets2008,Auguie2008,Rodriguez2011,Meinzer2014}. 

\begin{figure*}
\centering
	\includegraphics[width=15cm]{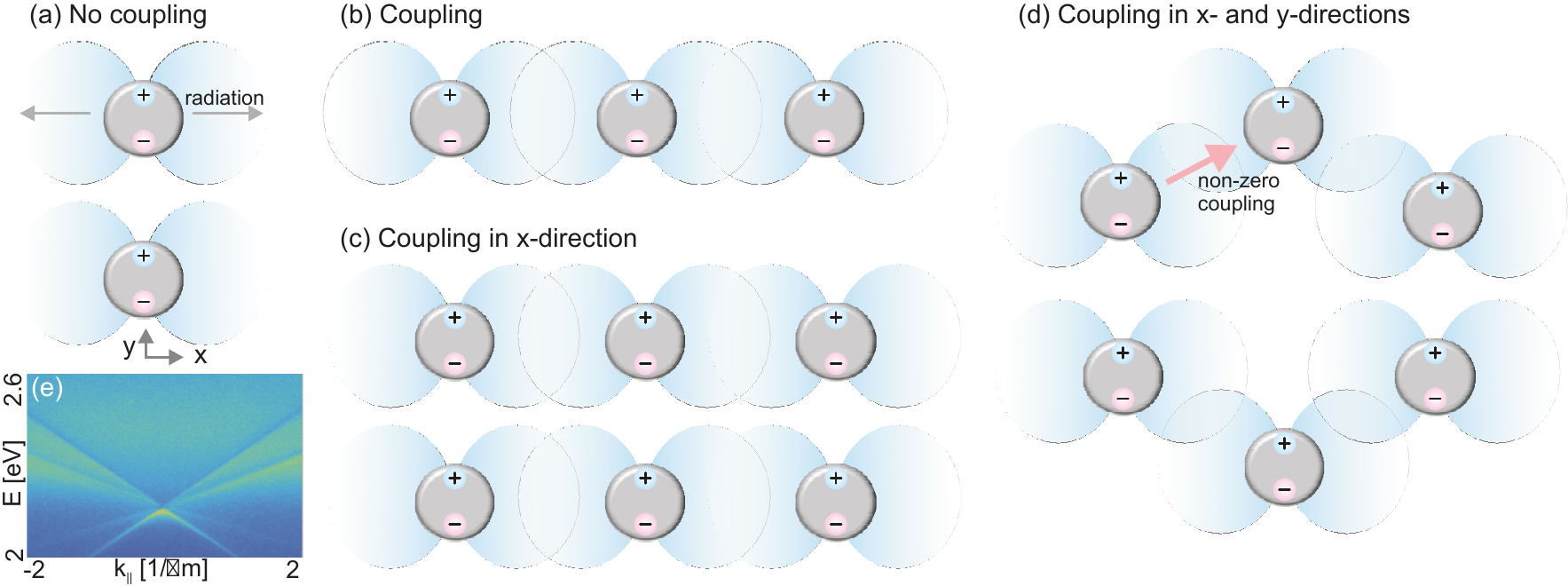}	
	\caption{Polarization dependence of the radiative couplings between the metallic nanoparticles. The radiative direction is orthogonal to polarization. Therefore if the metallic nanoparticles are placed parallel with the light polarization, there is no radiative coupling between them (a). If the nanoparticles form an array orthogonal to the polarization direction, they will couple with each other through the dipolar radiative interactions (b). Therefore, in rectangular arrays the SLRs depend only on one dimension (c). But in a more complex lattice, for example a honeycomb array (d), the radiative dipolar interactions would depend on both directions due to the non-zero interaction between the nearest particles along a combination of $x$ and $y$ directions, as the red arrow in (d) indicates. Thus the modes of the nanoparticle array are determined not only by the geometry of the lattice but also by its interplay with the polarization direction. We demonstrate the resulting rich mode structure in experiments on square, rectangular, hexagonal, honeycomb and Lieb nanoparticle geometries, and provide a simple approach to interpret and predict the observed dispersions, such as the one for a hexagonal lattice (e) (here $E$ is the light energy and $k_{//}$ the in-plane wave vector).}
	\label{fig:Fig1}
\end{figure*}

SLRs on plasmonic nanoparticle arrays show angle dependent dispersions and have significantly narrower linewidths compared with LSPRs on the individual particles. These features make metallic nanoparticle arrays suited for tailoring the light dispersion at the nanoscale. SLRs have been utilized in light harvesting \cite{Li2014}, emission control \cite{Vecchi2009,Rodriguez2012,Guo2015}, strong light-matter interaction \cite{Torma2015,Vakevainen2014,Shi2014}, and plasmonic lasing \cite{Stehr2003,Zhou2013,Meng2014,Yang2015,Hakala2016}. Recent works have also implemented SLRs in magneto-plasmonic responses in magnetic nanoparticle arrays \cite{Kataja2015,Maccaferri2015}, dark mode excitation in asymmetric dimer arrays \cite{Humphrey2016} and superlattice plasmons in hierarchical gold particle arrays \cite{Wang2015}. Even condensation phenomena have been theoretically studied \cite{Martikainen2014}. Yet another interesting aspect of SLRs stems from the fact that as the dipolar radiation pattern of the LSPR is non-isotropic, the effective radiative coupling in different lattice directions will depend strongly on polarization, as shown in Fig. \ref{fig:Fig1}. Moreover, since the coupling originates from the fairly slow-decaying radiation fields between the particles, any model relying only on the nearest-neighbor coupling, that is, the tight-binding model, is not sufficient to describe the system response. This raises the question whether such systems exhibit, for example, topologically non-trivial modes often found in tight-binding models \cite{Weick2013} and if so, whether these modes could have novel features. Recent progress in nanofabrication makes nanoparticle arrays with different lattice symmetries possible. However, more complex geometries have been experimentally investigated only under normal incident angle \cite{Humphrey2014}, thus providing no information on the system dispersion. Numerical models to calculate SLR dispersions, such as the discrete dipole approximation (DDA), have been provided \cite{Schatz2001,Kelly2003}, but an intuitive description from which one can straightforwardly determine the expected mode structure of also more complicated lattices has been missing. 

In this article, we explore the various SLRs supported by metallic nanoparticle arrays with different geometries. We provide a simple description of how complex lattice geometries affect the SLR dispersions with a given polarization. Our model uses simple diagrams to show how the different dispersions result from the coupling between the DO vectors of the lattice and the dipole orientations on the individual nanoparticles. Angle-resolved extinction spectra measured from silver nanoparticle arrays with square, hexagonal, rectangular, honeycomb and Lieb lattices are well explained by the simple model we provide for both TE- and TM-polarizations. Finite-difference time-domain (FDTD) simulations are also performed to verify our interpretation.

The complex dependence of the mode structure on the light polarization and the dipole orientations of individual particles suggests interesting possibilities if the polarization (the dipole orientation) is taken as a pseudospin degree of freedom. Our results show that the plasmonic nanoparticle arrays might be used for realizing novel types of spin-orbit coupling and thereby topological states of light, and provide an efficient approach for designing such systems. This prospect is discussed in the Conclusions.


\section{Model}

\begin{figure*}[!]
\centering
	\includegraphics[width=12cm]{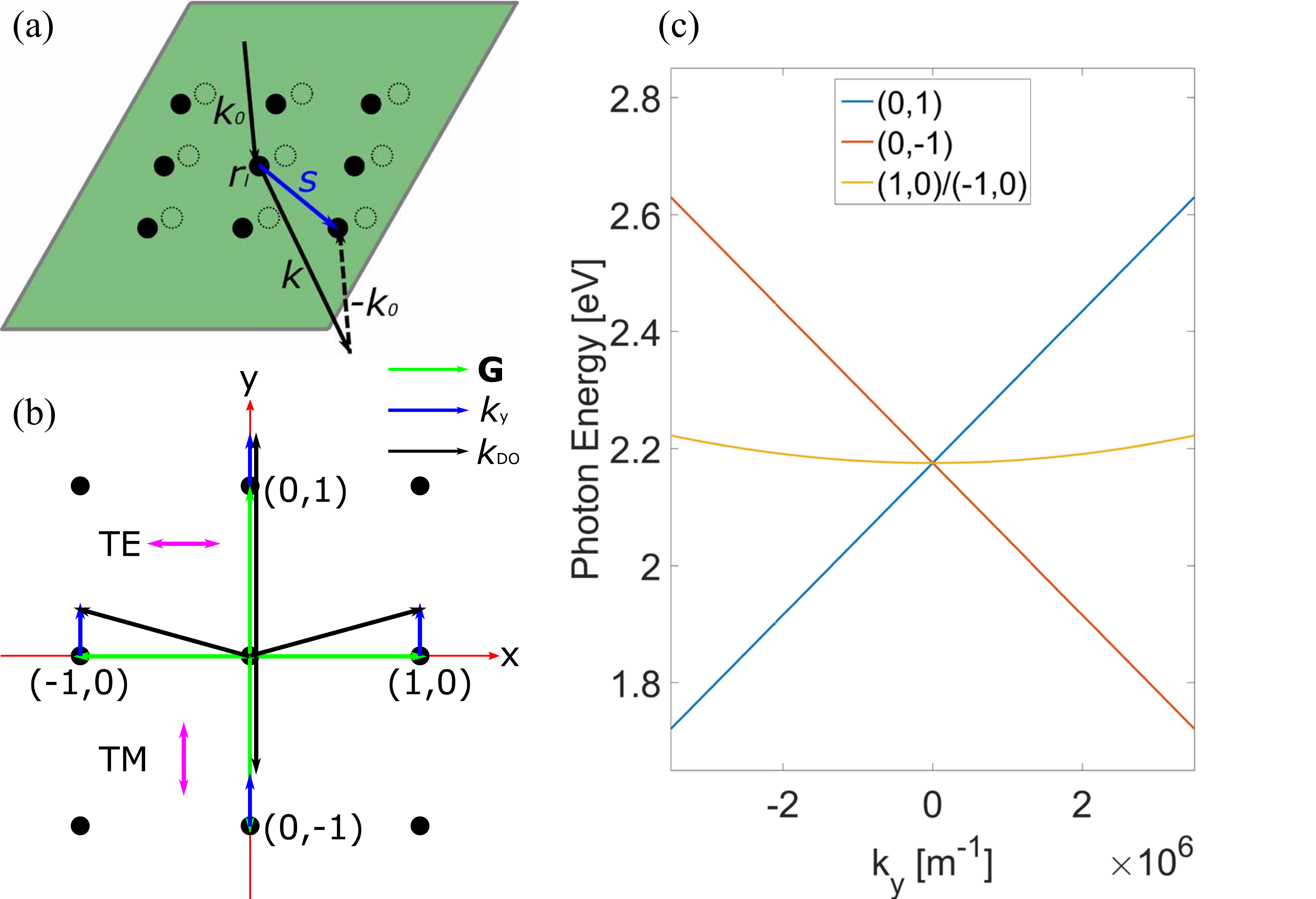}	
	\caption{(a) Scattering of a plane wave by a 2D particle array. Here \bm{$k_0$} and \bm{$k$} indicate the incoming and outgoing wave vectors, respectively, and $\bm{s}=\bm{k}-\bm{k_0}$ denotes the scattering vector. (b) The schematic of a square lattice in reciprocal space and the four lowest DOs (black arrows). The DOs $(0,1)$ and $(0,-1)$ have been slightly shifted horizontally to be better distinguished. The fuchsia arrows indicate the electric polarizations for TE- and TM-modes (TE and TM are defined with respect to $k_y$). (c) The calculated DOs $(0,1)$, $(0,-1)$ and the degenerate $(1,0)/(-1,0)$.}
	\label{fig:Fig2}
\end{figure*}

The SLRs involve a collectively scattered field that comprises components produced by scattering from all particles of the array \cite{Meinzer2014,Snoke2009}. A plane wave impinging on an array with a wave vector \bm{$k_0$} will be scattered by all the particles. The scattered wave can be also approximated by a plane wave \bm{$k$} at the far field limit. We define the scattering vector \bm{$s$} as the difference between \bm{$k_0$} and \bm{$k$} of each particle, as shown in Fig. \ref{fig:Fig2}(a). Then for a particle array, the amplitude of the total scattered wave will be proportional to $A_{sum}=\sum_{l}exp(i\bm{s}\cdot\bm{r_l})$ which is summed over the phase difference of all the particles. In the case of the Bravais lattice, where $\bm{r_l}=n_1\vec{a}_1+n_2\vec{a}_2$ and $\vec{a}_1$ and $\vec{a}_2$ are the primitive vectors of the lattice (here we consider 2D lattices), a non-zero total amplitude results only when $\bm{s}$ equals a reciprocal lattice vector $\bm{G}=m_1\vec{b}_1+m_2\vec{b}_2$ where $\vec{b}_1$ and $\vec{b}_2$ are the primitive vectors of the reciprocal lattice.

A more complex lattice may have a basis which consists of multiple particles in one unit cell, as the dotted circles show in Fig. \ref{fig:Fig2}(a). In such a case, the amplitude of the scattered light is the sum over all lattice sites and over each particle in the basis $A_{sum}=\sum_{l}\sum_{b}exp[i\bm{s}\cdot(\bm{r_l}+\bm{r_b})]=[\sum_{b}exp(i\bm{s}\cdot\bm{r_b})]\cdot[\sum_{l}exp(i\bm{s}\cdot\bm{r_l})]$. The sum $[\sum_{b}exp(i\bm{s}\cdot\bm{r_b})]$ affects the magnitude and phase of the scattering peaks. Therefore, we call it the envelope factor.

The dispersion relation for the DOs in a 2D periodic Bravais lattice is $\omega/c=|k_{//}+\bm{G}|$. If the lattice has a basis which consists of multiple particles in one unit cell, the dispersion maintains the same features, but the amplitude of the scattered light is modified by an envelope factor $[\sum_{b}exp(i\bm{G}\cdot\bm{r_b})]$, which is summed over the particles in one unit cell.

The expected SLR modes can be concluded from a simple diagram presenting the DOs and the polarization direction of the incident light. We take the square lattice as an example. Fig. \ref{fig:Fig2}(b) depicts the square lattice in reciprocal space, illustrating the four lowest order lattice vectors $\bm{G}$. The four lattice vectors have the same magnitude, meaning that the DOs $\omega/c=|k_y+\bm{G}|$ are degenerate when $k_y=0$, therefore the four branches have the same $\Gamma$-point. Considering the light incident with a small angle, there will be an in-plane wave vector, which here we assume is $k_y$, added to the DOs. In such a case, the orders $(1,0)$ and $(-1,0)$ still have the same magnitude so they maintain the degeneracy, as shown by the black arrows $\bm{k}_{\text{DO}}$ in Fig. \ref{fig:Fig2}(b). But the DO vectors $(0,1)$ and $(0,-1)$ then have different magnitudes so they become two different branches, linearly dependent on $k_y$. By this means we can determine the DOs supported by a square lattice, as shown in Fig. \ref{fig:Fig2}(c). There are two linearly dispersed branches that correspond to the DOs $(0,1)$ and $(0,-1)$, and one hyperbolically dispersed branch corresponds to the degenerate DOs $(1,0)$ and $(-1,0)$. 

Besides magnitude, the directions of the DO vectors $\bm{k}_{\text{DO}}$ describe also their propagation directions at a certain wavelength and incident angle. For metallic nanoparticle arrays, the supported plasmonic modes are not only dependent on the geometrical properties but also on the excitation polarization. Let us consider two orthogonal polarization directions, TE and TM, as depicted by the fuchsia arrows in Fig. \ref{fig:Fig2}(b). A nanoparticle driven by an incident field of a certain polarization mainly radiates in the orthogonal direction and couples well with modes whose DO vectors $\bm{k}_{\text{DO}}$ are in that direction. It does not excite a mode whose $\bm{k}_{\text{DO}}$ is parallel to the polarization, see the schematics in Fig. \ref{fig:Fig1}. Therefore we expect to excite the DOs $(0,1)$ and $(0,-1)$ only by TE-polarized light, and the degenerate DOs $(1,0)$ and $(-1,0)$ preferably by TM-polarized light, in the case of a small angle of incidence.

In the following, we show that this approach efficiently describes the modes of also more complex lattices and allows a straightforward interpretation of the experimentally observed dispersions.


\section{Sample fabrication and optical characterization}

\begin{figure*}[!]
\centering
	\includegraphics[width=14cm]{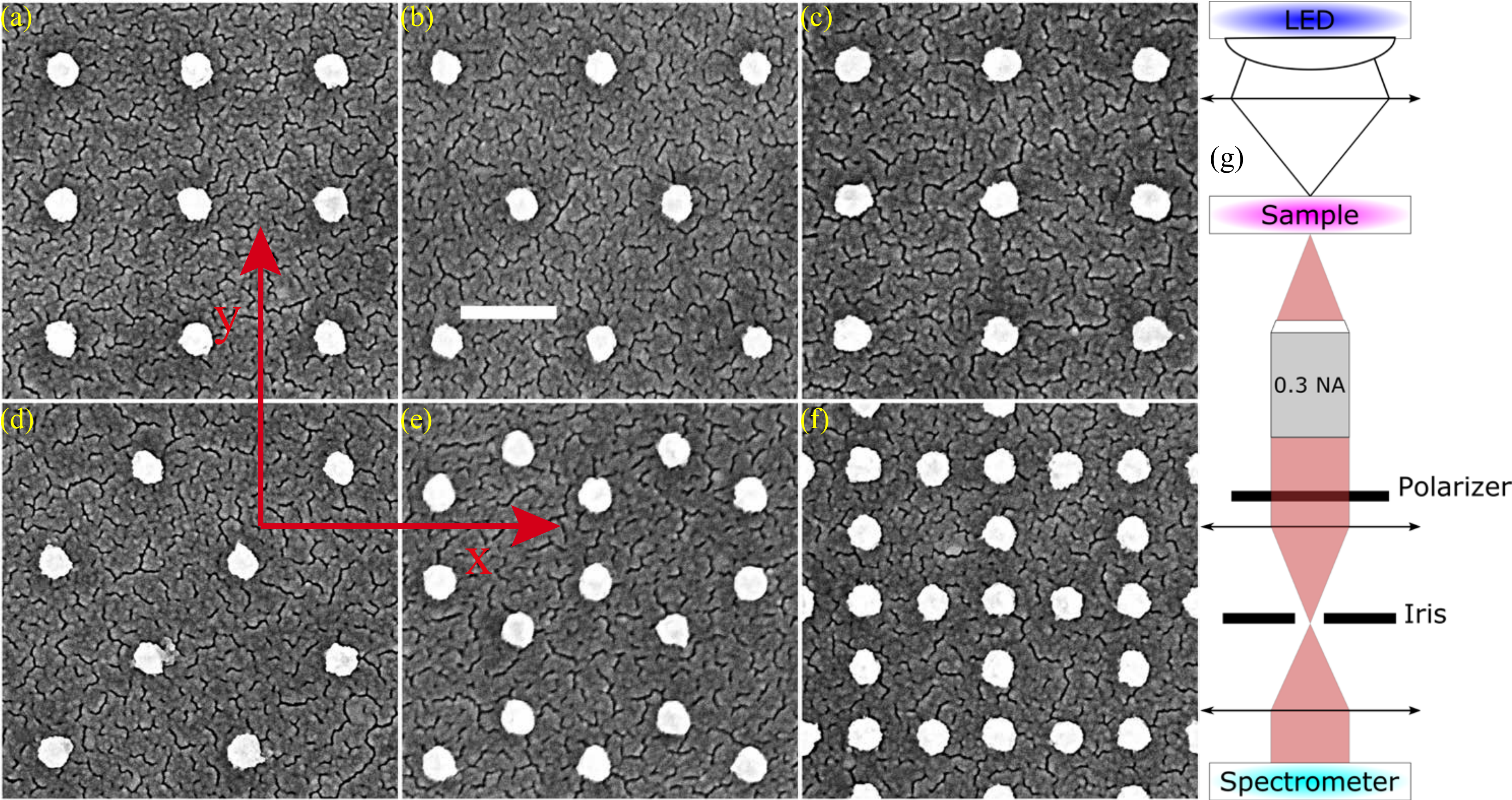}
	\caption{(a)-(f) Scanning electron micrographs (SEMs) of square, hexagonal, rectangular, 45-degree rotated square, honeycomb and Lieb lattice arrays. The scale bar in (b) is $300 \text{ nm}$ and the same for all. (g) The measurement setup. The angle-resolved transmission spectra are collected by focusing the image of the back focal plane of the objective to the entrance slit of the spectrometer. The entrance slit is parallel to the $y$-axis of the sample. For transmission measurements, a white LED light source is used.}
	\label{fig:Fig3}
\end{figure*}

We fabricate silver nanoparticle arrays on borosilicate glass with different patterns by e-beam lithography. The designed patterns include square, hexagonal, rectangular, $45$-degree rotated square, honeycomb and Lieb lattices, as shown in the top-view scanning electron micrographs (SEMs) of the fabricated samples in Figs. \ref{fig:Fig3}(a)-(f). Each array has a size of $100\text{ {\micro}m} \times 100\text{ {\micro}m} $, and the silver nanoparticles have a height of $30 \text{ nm}$ and a diameter of $60 \text{ nm}$, with $2 \text{ nm}$ titanium below as an adhesive layer. The distances between the particles are chosen such that the diffractive edges ($\Gamma$-points) of all structures are at the same energy, which will be explained in detail below.

Angle-resolved transmission spectra $T=I_{\text{structure}}/I_{\text{reference}}$ are measured by focusing the image of the back focal plane of the objective to the entrance slit of a spectrometer, as shown in the schematic Fig. \ref{fig:Fig3}(g). A white LED is used as the light source, and a polarizer is placed before the spectrometer to control the polarization of the detected light. The sample slide is embedded by an index-matching oil (refractive index $n=1.52$) and covered by a borosilicate superstrate to provide a symmetric optical environment for the arrays. The extinction spectra are then obtained by $(1-T)$ and subsequently used for calculating the dispersions on each array.

\begin{figure*}[!]
\centering
	\includegraphics[width=15cm]{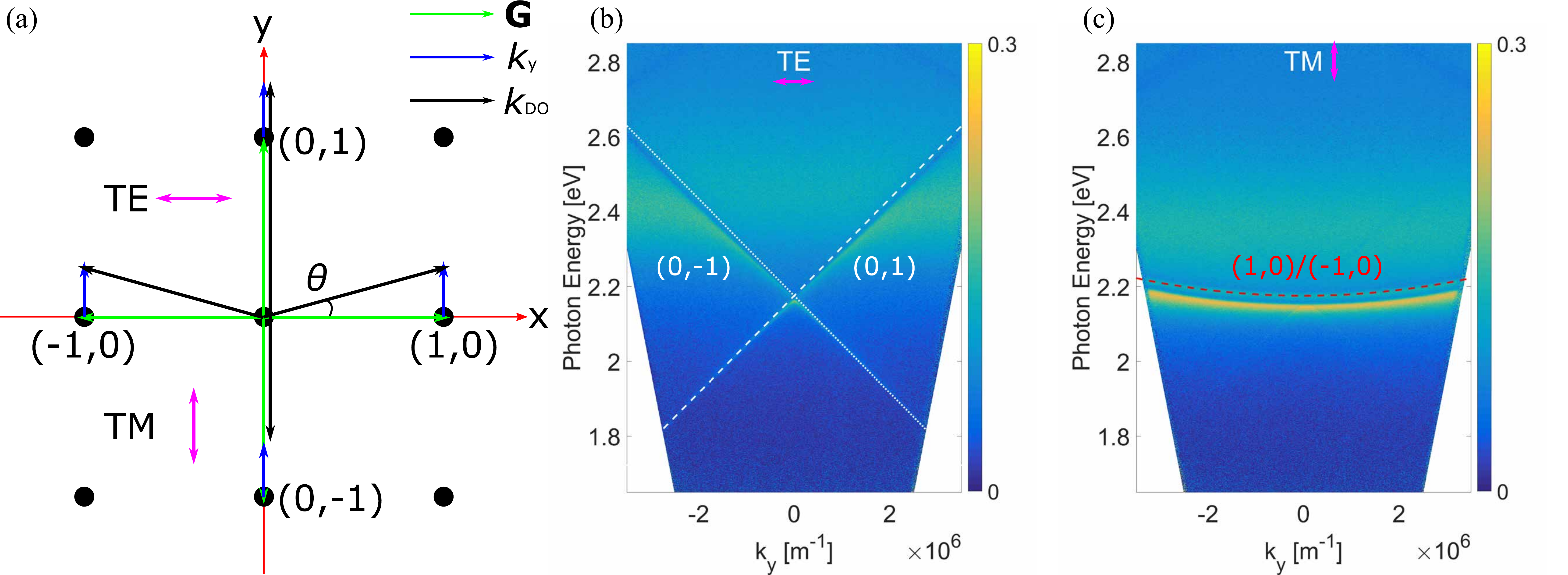}	
	\caption{(a) The schematic of a square lattice in reciprocal space and the four lowest DOs (black arrows). The DOs $(0,1)$ and $(0,-1)$ have been slightly shifted horizontally to be better distinguished. The fuchsia arrows indicate the electric polarizations for TE- and TM-modes (TE and TM are defined with respect to $k_y$). The measured extinction spectra for (b) TE- and (c) TM-polarized lights. Dashed and dotted white lines represent the calculated DOs $(0,1)$ and $(0, -1)$, respectively. The dashed red line represents the calculated degenerate DOs $(1,0)$ and $(-1,0)$. The broad feature around $2.4 \text{ eV}$ is the localized single particle resonance (LSPR), here and in Figs. \ref{fig:Fig5}-\ref{fig:Fig9}.}
	\label{fig:Fig4}
\end{figure*}

\section{Measurement results}

\subsection{Square lattice}

Figs. \ref{fig:Fig4}(b) and (c) show the extinction of a square lattice with a periodicity of $p_x=p_y=375 \text{ nm}$ for TE- and TM-polarizations, respectively. Dispersions as a function of the in-plane wave vectors in the $y$ direction, $k_y$, are considered. TE-polarization means the electrical field oscillates perpendicular to the in-plane wave vector $k_y$, and TM parallel to it. The different dispersions among the TE- and TM-modes originate from the different mode vectors of the lowest DOs on the square lattice. 

\begin{figure*}[ht!]
\centering
	\includegraphics[width=15cm]{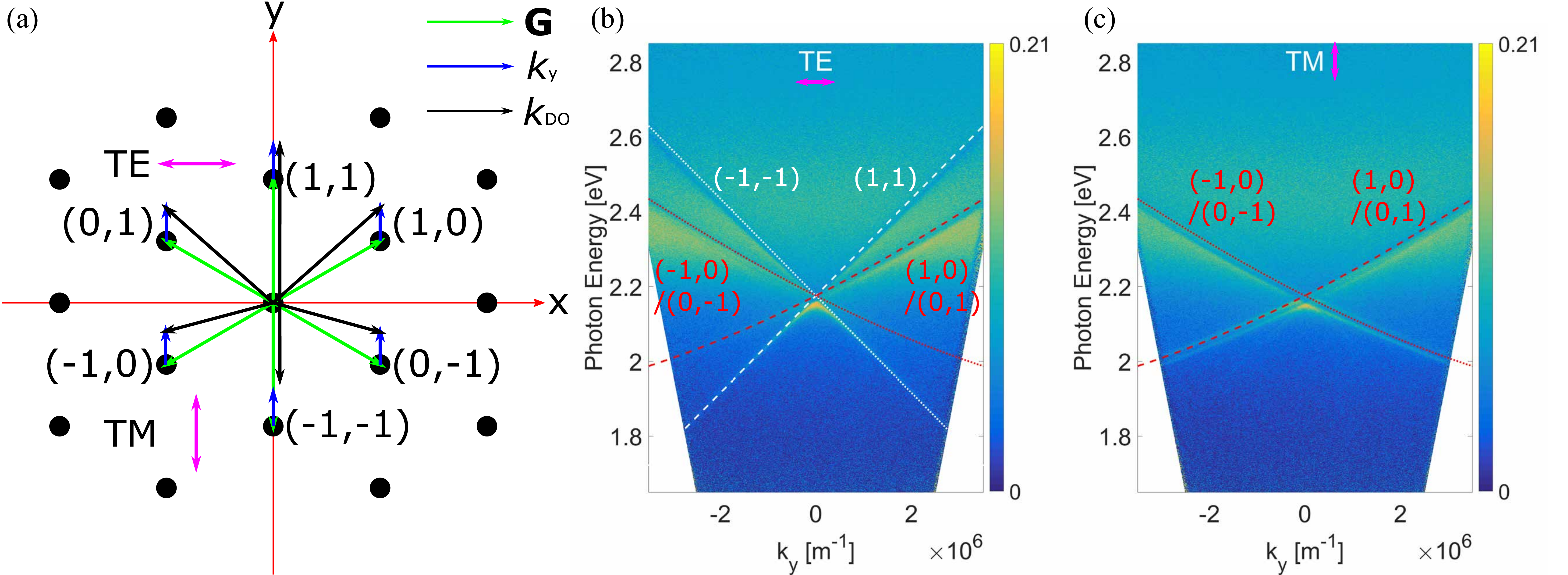}	
	\caption{(a) The schematic of a hexagonal lattice in reciprocal space and the six lowest DOs. Measured extinction and calculated DOs for (b) TE- and (c) TM-polarized light. Dashed and dotted white lines represent the DOs $(1,1)$ and $(-1,-1)$, respectively. Dashed and dotted red lines represent the degenerate DOs $(1,0)/(0,1)$ and $(0,-1)/(-1,0)$, respectively.}
	\label{fig:Fig5}
\end{figure*}

As explained in Section II, for TE-polarized incident light, $(0,1)$ and $(0,-1)$ DOs are excited and linearly dispersed with the in-plane wave vector. However, (1,0) and (-1,0) DOs are extremely weakly coupled to TE-polarized light because they propagate nearly-orthogonally to the radiative direction of the electric dipole. Therefore they are not visible in Fig. \ref{fig:Fig4}(b). By increasing $\bm{k}_{\text{DO}}$, one could make these modes less orthogonal to the radiative direction and thus visible with TE-polarized light, but in our measurement setup, the maximum detectable $k_y$ at visible range is four times smaller than the magnitude of the lowest lattice vector $|\bm{G}|$, thus in the schematic Fig. \ref{fig:Fig4}(a) the maximum angle $\theta$ obtainable for us at present is $13^{\circ}$. On the other hand, a TM-polarized incident light can excite only the degenerate $(1,0)$ and $(-1,0)$ modes since $(0,1)$ and $(0,-1)$ DOs both propagate orthogonally to the radiative direction and are therefore suppressed: the degenerate $(1,0)/(-1,0)$ modes are shown in Fig. \ref{fig:Fig4}(c).

In Figs. \ref{fig:Fig4}(b) and (c) we plot also the corresponding $(0,1)$, $(0,-1)$ and $(1,0)/(-1,0)$ DOs from which we can see that the SLRs follow the calculated DOs quite well. The red-shifts of the SLRs are due to the coupling to LSPRs.

\subsection{Hexagonal lattice}

Extinctions of a hexagonal lattice with a nanoparticle separation of $433 \text{ nm}$ show more SLRs than the square lattice case for both TE- and TM-modes, as Figs. \ref{fig:Fig5}(b) and (c) indicate. These extra branches are due to the complex structure of the lattice in reciprocal space: a hexagonal lattice has six lowest order lattice vectors of equal length, as shown in Fig. \ref{fig:Fig5}(a). The six lowest DOs then have the same $\Gamma$-point. With an in-plane wave vector $k_y$, the DOs $(1,0)$ and $(0,1)$ will stay degenerate since they still have the same magnitude. The same holds for the DOs $(0,-1)$ and $(-1,0)$ so they are also degenerate in the dispersion diagram. The DOs $(1,1)$ and $(-1,-1)$ are not degenerate. TE-polarized light can now excite all the modes; taking into consideration the degeneracies, this leads to four modes visible in the experiment with TE-polarized incident field, as shown in Fig. \ref{fig:Fig5}(b).

The DOs $(1,1)$ and $(-1,-1)$ both propagate parallel to the $y$-axis, so when the excitation field is TM-polarized, the propagating direction is orthogonal to the radiative direction of the electric dipole on each nanoparticle. These two branches are therefore no longer visible. This explains why there are only two branches of SLRs in the extinction spectra for TM-polarized light, as shown in Fig. \ref{fig:Fig5}(c).

\subsection{Rectangular lattice}

\begin{figure*}
\centering
	\includegraphics[width=15cm]{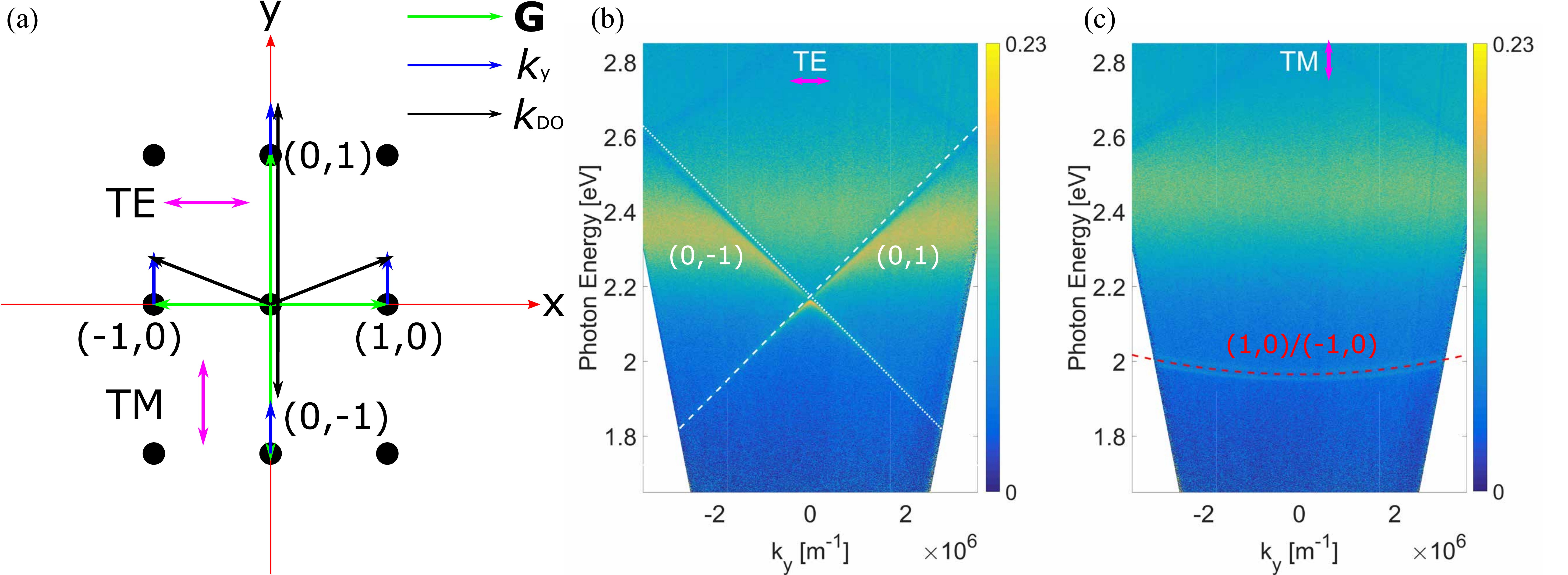}	
	\caption{(a) The schematic of a rectangular lattice in reciprocal space and the four lowest DOs. For clarity of the schematic, the distances along $x$- and $y$-axes in the reciprocal lattice are not in scale with those in the real sample. Measured extinction and calculated DOs for (b) TE- and (c) TM-polarized light. Dashed and dotted white lines represent the DOs $(0,1)$ and $(0, -1)$, respectively. The dashed red line represents the degenerate DOs $(1,0)/(-1,0)$.}
	\label{fig:Fig6}
\end{figure*}

\begin{figure*}
\centering
	\includegraphics[width=15cm]{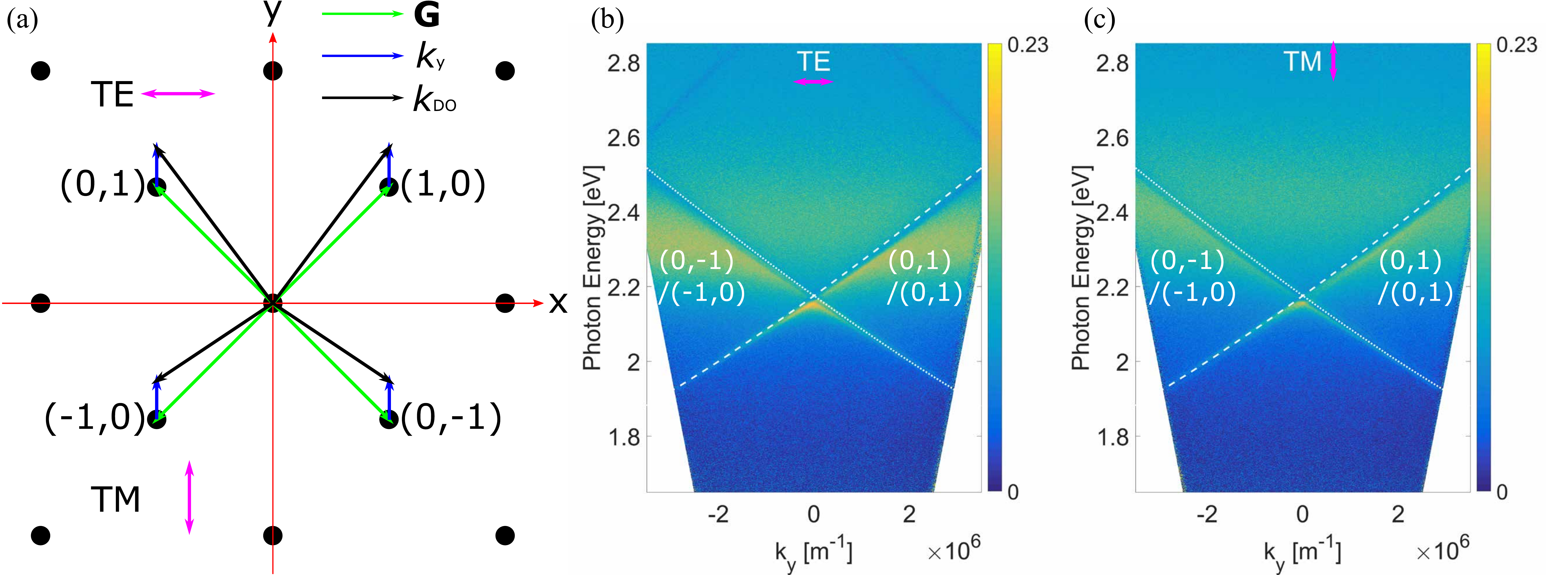}	
	\caption{(a) The schematic of a $45$-degree rotated square lattice in reciprocal space and the four lowest DOs. Measured extinction and calculated DOs for (b) TE- and (c) TM-polarized light. Dashed and dotted white lines represent the degenerate DOs $(1,0)/(0,1)$ and $(-1,0)/(0,-1)$, respectively.}
	\label{fig:Fig7}
\end{figure*}

The lower symmetry in a rectangular lattice $(p_x \neq p_y)$ compared to the square one has a strong impact on the supported SLRs. The two primitive vectors in reciprocal space have different magnitudes so the $\Gamma$-points of the DOs split in energy, as shown in Fig. \ref{fig:Fig6}(a). As the square lattice, the $(0,1)$ and $(0,-1)$ DOs are supported and linearly dispersed with the in-plane wave vector for the TE-polarized excitation field, and $(1,0)$ and $(-1,0)$ DOs are degenerate and hyperbolically dispersed with the in-plane wave vector for the TM-polarized excitation field. However, the TE- and TM-modes have different $\Gamma$-points due to the periodicity mismatch, as shown in the measured extinctions in Figs. \ref{fig:Fig6}(b) and (c) of a rectangular array where the periodicities along $x$- and $y$-axes are $415 \text{ nm}$ and $375 \text{ nm}$, respectively.

Such $\Gamma$-point inequality has been reported in \cite{Humphrey2014}, where the authors have measured the zero-angle extinction spectra for rectangular lattice arrays and found different peak positions for different polarizations, noticing that the important distance is the particle separation in the direction perpendicular to the incident electric field. Here, we point out the origin of this phenomenon: it naturally arises from the general framework we propose for determining and quantifying the SLR modes in different lattice geometries.

\subsection{45-Degree rotated square lattice}

We fabricate also arrays with a square lattice that is rotated 45 degrees with respect to the $x$- and $y$-axes (a rhombus with right angles), with the particle separation the same as the square lattice ($375 \text{ nm}$). Then neither of its two reciprocal primitive vectors are parallel to the $x$-/$y$-axis, as shown in Fig. \ref{fig:Fig7}(a). In such a case, the four lowest DOs all have the same $\Gamma$-point position. With an in-plane wave vector $k_y$, DOs $(1,0)$ and $(0,1)$ are degenerate since they always have the same magnitude. Similarly, the DOs $(0,-1)$ and $(-1,0)$ are degenerate.

Additionally, all the lowest DOs propagate nearly-diagonally with increasing $k_y$ –-- $(1,0)/(0,1)$ shifting towards the $y$-axis and $(0,-1)/(-1,0)$ shifting towards the $x$-axis, all with small angles (in our measurements) deviating from the diagonal direction. Therefore both TE- and TM-polarized field can excite these branches simultaneously. As Figs. \ref{fig:Fig7}(b) and (c) show, the calculated lowest DOs match well with the measured extinctions, meaning that the SLRs become polarization-independent for this lattice type. However, the second-lowest DOs are then either parallel to the $x$- or $y$-axis, so they become polarization dependent. Therefore in the TE-polarization extinction, there are two branches in the high energy regime which correspond to the second-lowest DOs $(1,1)$ and $(-1,-1)$, see the faint features in the upper part of Fig. \ref{fig:Fig7}(b). But in TM-polarization extinction, no such features have been observed since the $(1,-1)/(-1,1)$ modes are hyperbolically dispersed and non-detectable in the measured wavelength regime.

\begin{figure*}[!]
\centering
	\includegraphics[width=15cm]{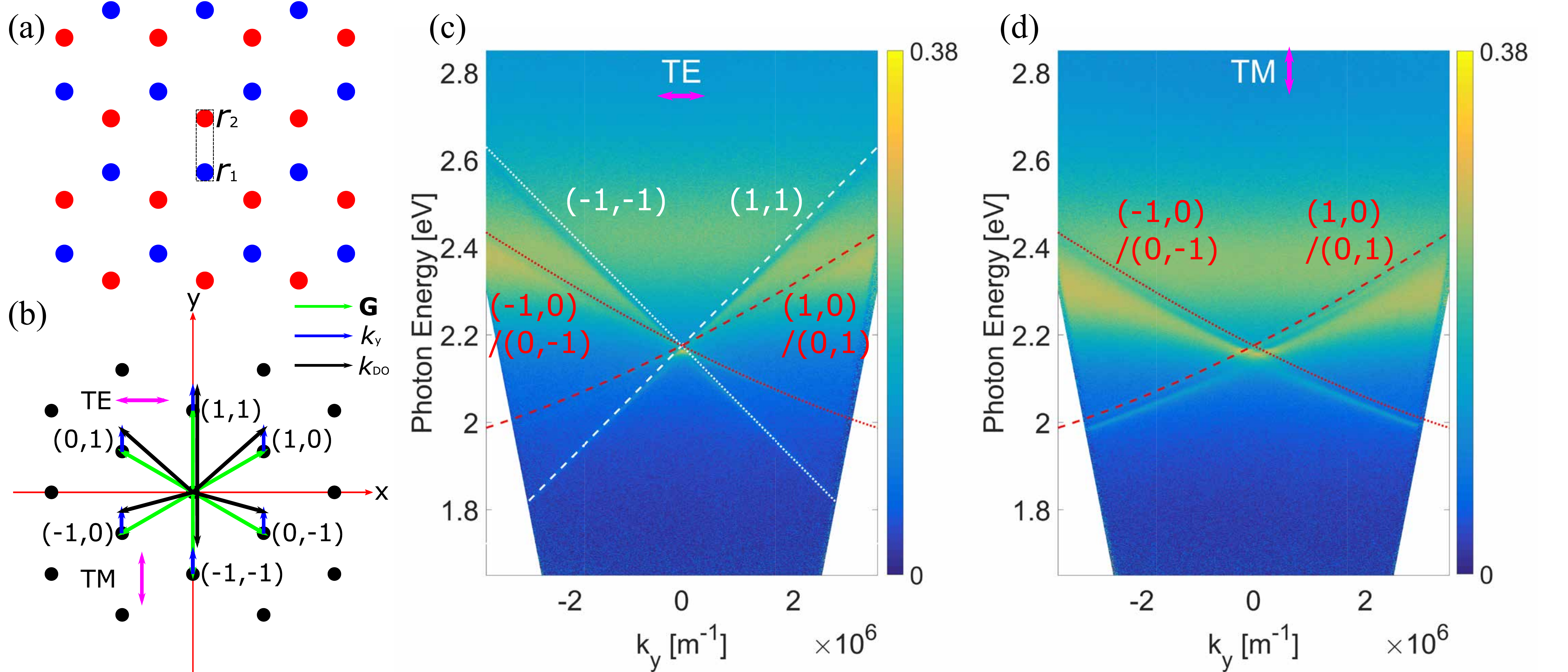}	
	\caption{The schematics of (a) a honeycomb lattice, with its unit cells that constitute a hexagonal lattice, and (b) a hexagonal lattice in reciprocal space and the six lowest DOs. Measured extinction and calculated DOs for (c) TE- and (d) TM-polarized light. Dashed and dotted white lines represent the DOs $(1,1)$ and $(-1,-1)$, respectively. Dashed and dotted red lines represent the degenerate DOs $(1,0)/(0,1)$ and $(0,-1)/(-1,0)$, respectively.}
	\label{fig:Fig8}
\end{figure*}

\subsection{Honeycomb lattice}

We consider also more complex structures, namely geometries composed by a Bravais lattice with a basis, that is, a unit cell with several sites. As mentioned in Section II, then there will be a contribution from the scattering of each particle within one unit cell to the overall amplitude as an envelope factor $[\sum_{b}exp(i\bm{G}\cdot\bm{r_b})]$ which depends on the corresponding reciprocal lattice vector and the particle locations within the unit cell. Since the envelope factor depends on $\bm{G}$, it can be different for different DOs.

As shown in Fig. \ref{fig:Fig8}(a), a honeycomb structure is constituted of a hexagonal lattice with two particles within each unit cell. The envelope factors from these two particles to the six lowest DOs are $1/2 + \sqrt{3}/2i$ for $(1,0)$, $(0,1)$, $(-1,-1)$ and $1/2 - \sqrt{3}/2i$ for $(1,1)$, $(-1,0)$, $(0,-1)$, respectively. Both factors have unity magnitude, but opposite argument angles, meaning that the DOs from a honeycomb array have different phase shifts compared with a hexagonal array. For a certain branch, the phase shift depends only on the particle locations within one unit cell and thus can be tuned easily by displacing the particles within the unit cell. 

Figs. \ref{fig:Fig8}(c) and (d) show the measured extinctions for a honeycomb array with a nearest particle separation of $250 \text{ nm}$, and the calculated DOs for its corresponding Bravais hexagonal lattice. The measured SLRs follow well with the calculated DOs. The only prominent difference between the extinction of the honeycomb array and its corresponding hexagonal array, as shown in Figs. \ref{fig:Fig5}(b) and (c), is that the honeycomb array has a higher extinction. The increasing extinction is due to a larger number of particles within one unit cell, resulting in a higher filling fraction in a honeycomb structure. Another difference is the complex envelope factors of the DOs in honeycomb lattice, which in a hexagonal lattice is a constant unity. The phase shift originating from the envelope factors is not expected to be reflected in extinction, however, it can be interesting when using such lattices for designing beams, e.g. in nanoscale lasing.

\subsection{Lieb lattice}

\begin{figure*}
\centering
	\includegraphics[width=15cm]{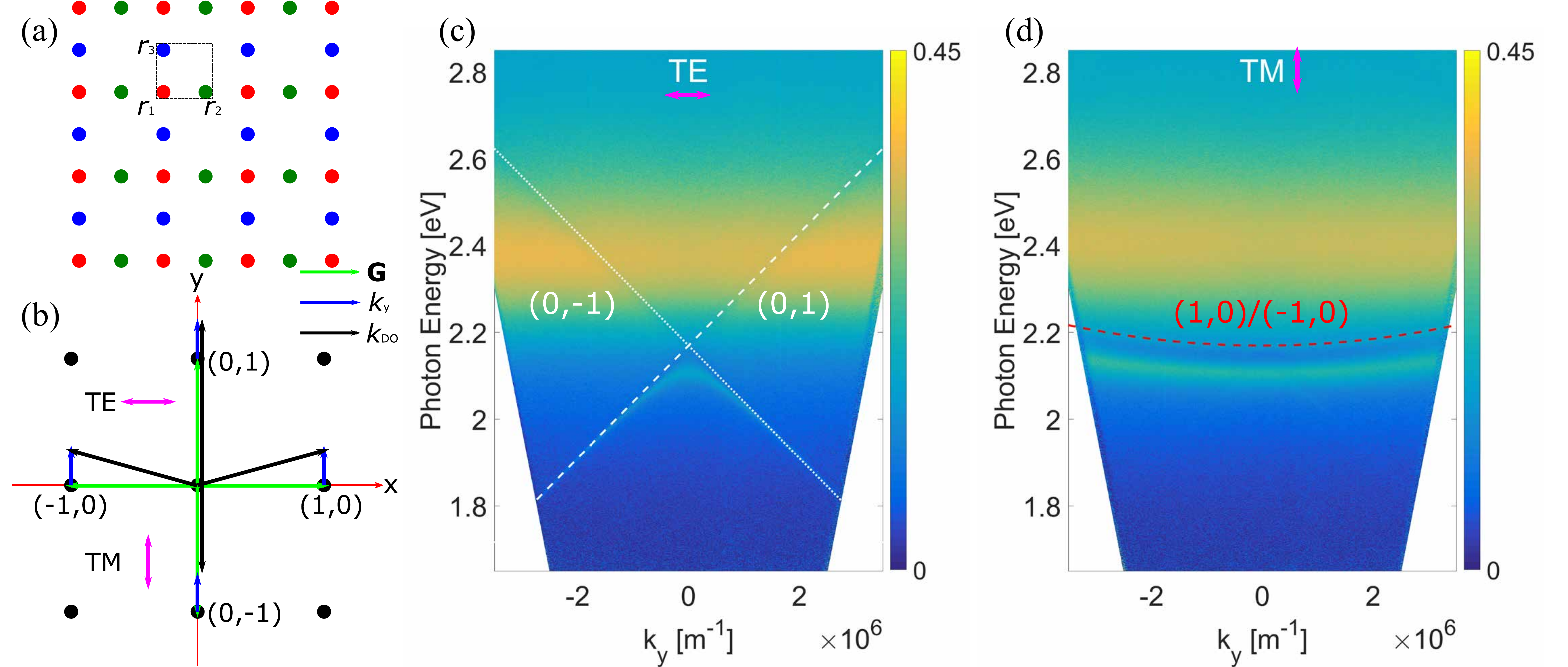}	
	\caption{The schematics of (a) a Lieb lattice, with its unit cells that constitute a square lattice, and (b) a square lattice in reciprocal space and the four lowest DOs. Measured extinction and calculated DOs for (c) TE- and (d) TM-polarized light. Dashed and dotted white lines represent the DOs $(0,1)$ and $(0,-1)$, respectively. The dashed red line represents the degenerate DOs $(1,0)/(-1,0)$.}
	\label{fig:Fig9}
\end{figure*}

Similar to the honeycomb case, we can calculate the DOs of a Lieb lattice, which is constituted by a square lattice with three particles within each unit cell, as shown in Fig. \ref{fig:Fig9}(a). The lowest DOs are then the same as for a square lattice and their envelope factors from the three particles are all one. The lowest orders of diffracted light from a Lieb array are therefore the same as from a square array, without any phase shifts.

\begin{figure*}
\centering
	\includegraphics[width=16cm]{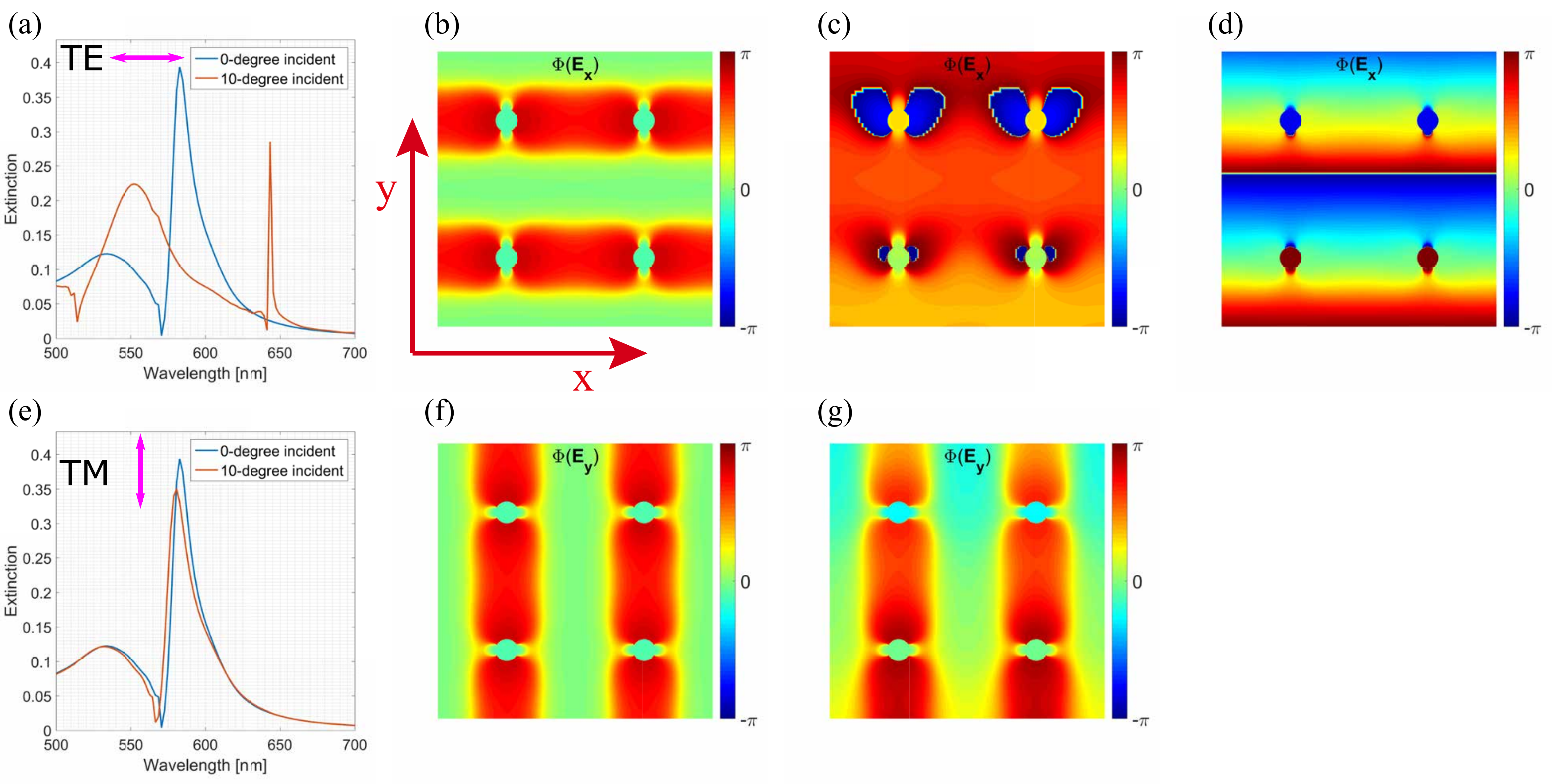}	
	\caption{FDTD simulated extinction spectra for a square lattice under normal and $10$-degree incidence with (a) TE- and (e) TM-polarized light. The field distributions $\Phi(E_x)$ for TE-polarized light at $583 \text{ nm}$ wavelength under normal incidence (b), under $10$-degree incidence at $553 \text{ nm}$ (c) and $643 \text{ nm}$ (d) wavelengths. For TM-polarized light, the field distributions $\Phi(E_y)$ at $583 \text{ nm}$ wavelength under normal incidence (f) and under $10$-degree incidence at $581 \text{ nm}$ wavelength (g). Here (b) and (f) relate to the SLRs corresponding to the degenerate DOs at the $\Gamma$-point in Figs. \ref{fig:Fig4}(b) and \ref{fig:Fig4}(c), respectively. (c) corresponds to the $(0,1)$ DO and (d) to the $(0,-1)$ DO modes in Fig. \ref{fig:Fig4}(b). (g) relates to the degenerate modes of DOs $(1,0)/(-1,0)$ of Fig. \ref{fig:Fig4}(c) slightly away from the $\Gamma$-point.}
	\label{fig:Fig10}
\end{figure*}

Figs. \ref{fig:Fig9}(c) and (d) show the measured extinctions of a Lieb array with a nearest particle separation $188 \text{ nm}$ and the calculated lowest DOs for the corresponding Bravais square lattice. The similarity between the SLRs of the Lieb lattice and of the square array, as shown in Figs. \ref{fig:Fig4}(b) and (c), is not as big as those between the honeycomb array and the hexagonal array. A three-fold increase in filling fraction in the Lieb lattice enhances the coupling strength between the LSPRs and DOs, resulting in a stronger modification of the SLR energies even though the DOs are exactly the same. This can be seen from the higher extinction of the Lieb array and the more curved-dispersed TE-polarized SLRs near the $\Gamma$-point. Note that the dispersions are very different from the photonic Lieb lattices with nearest neighbor hopping \cite{Vicencio2015,Mukherjee2015,Baboux2016,Diebel2016}.

\section{Numerical simulations}

From our lattice geometry arguments and measurement results, we see that the supported SLRs of arrays with different geometries are mostly dependent on their DOs and the orientations of in-plane electric dipoles of individual nanoparticles. In other words, the far-field properties of a specific lattice structure (the DOs) determine the coupling directions in which a specific near-field profile (dipole orientation) oscillates in phase. Conversely, the radiative direction of the in-plane electric dipoles determines whether or not a specific DO can be excited.

To further confirm, at the microscopic scale, the general framework given by geometry arguments and the measurements, we perform FDTD simulations with commercial software (FDTD solutions, Lumerical Inc.) for square arrays with the same periodicity in Section IV as an example. Fig. \ref{fig:Fig10}(a) shows the simulated extinction spectra with TE-polarized light under normal and $10$-degree incident angles. The distribution map of the in-plane electric field component phase, which for TE-polarization is $\Phi(E_x)$, shows that the electric dipoles on individual nanoparticles oscillate in phase with each other, forming a standing wave, as indicated in Fig. \ref{fig:Fig10}(b). For a 10-degree incident angle, as shown in Figs. \ref{fig:Fig10}(c) and (d), the dipole moments of individual nanoparticles no longer oscillate in phase with each other, but form a phase front along the $y$-axis. 

The simulated extinction spectra of TM-polarized light in Fig. \ref{fig:Fig10}(e) show exactly the same resonance wavelength with TE-polarization under normal incidence (at $\Gamma$-point) but a slightly different resonance wavelength under $10$-degree incident angle, due to the low-dispersed property of TM-mode. The distribution map of the in-plane electric field component phase $\Phi(E_y)$ in Fig.\ref{fig:Fig10}(f) shows that under normal incident angle, the electric dipoles on individual nanoparticles oscillate in phase with each other, forming a standing wave along the $x$-axis. However under a $10$-degree incident angle, as shown in Fig. \ref{fig:Fig10}(g), the dipole moments of individual nanoparticles no longer oscillate in phase with each other and form a phase front nearly along the $x$-axis. The different wave fronts of TE- and TM-polarization cases correspond to the DO vector directions of the SLR modes that are excited. The DO vectors determine the directions in which the specific modes are propagating.

\section{Conclusions}

We demonstrate experimentally a rich variety of dispersions of distinctive SLRs supported by silver nanoparticle arrays of different geometries. Square, hexagonal, rectangular, honeycomb and Lieb arrays show remarkably different and polarization-dependent extinction dispersions, while a $45$-degree rotated square array shows less sensitivity to the choice of the TE- or TM-polarized incident light. While previous work has studied the case of normal incidence \cite{Humphrey2014}, the present work constitutes the first systematic study of SLR dispersions for arbitrary incident angles and for various lattice geometries.

We propose an efficient generic model to explain and predict the features of different plasmonic lattice geometries. Using simple diagrams, we show how the DO vectors of the corresponding Bravais lattice and the in-plane electric dipole orientation determine the SLRs supported by certain geometrical structures. In a more complex structure, the particles within one unit cell contribute to an envelope factor for each DO. The modes determined in this way have excellent agreement with the measured spectra. Furthermore, this model also reveals the principal propagation direction of each mode, along which the mode maintains its coherence properties. This has been verified by numerical simulations of the near-field distributions in a square lattice. 

The radiation fields of individual particles extend over several unit cells of the structure, making plasmonic lattices an excellent  platform for studying physics beyond the nearest neighbor hopping regime. Furthermore, the role of disorder can be quite different in radiatively coupled plasmonic lattices compared to photonic or plasmonic lattices with evanescent nearest neighbor coupling. Totally new areas of research may open by interpreting the individual particle dipole orientation as a pseudospin degree of freedom, or more generally, as a two-level or two-band system. The Hamiltonian describing any two-level system has the generic form $H=\epsilon(\bm{k})I_{2\times2}+\bm{d}(\bm{k})\cdot\bm{\sigma}$, where $\bm{\sigma}$ is a vector of the Pauli matrices, and $\bm{d}(\bm{k})$ determines the types of couplings and $\bm{k}$ is a parameter. This Hamiltonian describes, for instance, graphene and spin-orbit coupled systems, and with suitable couplings and symmetries, Dirac points, gap openings and topological phases can be found \cite{Bernevig2013,Hasan2010,Lu2014,Lu2016,Haldane2008,Raghu2008}. As we have shown, in plasmonic lattices, the dispersion depends on the polarization, i.e., on the particle dipole orientation. This can be viewed analogously to having a non-trivial $\sigma_z$ term (here $z$ now refers to the dipole orientation pseudospin, not to the spatial coordinates of the lattice). The spin-orbit coupling terms proportional to $\sigma_x$ and $\sigma_y$ are not present here. However, we propose that they can be introduced by utilizing non-trivial particle shapes, or magnetic nanoparticles \cite{Kataja2015,Maccaferri2015} where the two polarization (dipole orientation) directions are coupled due to intrinsic spin-orbit coupling in the magnetic material, and such couplings could be made momentum($\bm{k}$) dependent by designing the lattice geometry. It is especially interesting to envision that such non-trivial lattice systems may lead to new types of nanoscale lasing phenomena, so far observed only in simple square or rectangular lattices.

\begin{acknowledgments}


This work was supported by the Academy of Finland through its Centres of Excellence Programme (Project No. 284621, No. 263347 and No. 272490) and by the European Research Council (Grant No. ERC-2013-AdG-340748-CODE). This  article  is  based  on work  from  COST  Action  MP1403  Nanoscale  Quantum Optics,  supported  by  COST  (European  Cooperation  in  Science  and  Technology). Part of the research was performed at the Micronova Nanofabrication Centre, supported by Aalto University.

\end{acknowledgments}



\newcommand{\noop}[1]{}

\end{document}